\documentclass[rnote]{aa} 

\usepackage{graphicx}
\usepackage{txfonts}
\usepackage{natbib}
\usepackage{longtable}

\begin{document}

\title{Estimation of  high-resolution dust column density maps}
\subtitle{Empirical model fits}

\author{M.     Juvela\inst{1},
J.     Montillaud\inst{1}
}

\institute{
Department of Physics, P.O.Box 64, FI-00014, University of Helsinki,
Finland, {\em mika.juvela@helsinki.fi}
}

\authorrunning{M. Juvela et al.}

\date{Received September 15, 1996; accepted March 16, 1997}

\abstract
{
Sub-millimetre dust emission is an important tracer of density $N$ of
dense interstellar clouds. One has to combine surface brightness
information at different spatial resolutions, and specific methods are
needed to derive $N$ at a resolution higher than the lowest resolution
of the observations. Some methods have been discussed in the
literature, including a method (in the following, method B) that
constructs the $N$ estimate in stages, where the smallest spatial scales
being derived only use the shortest wavelength maps.}
{
We propose simple model fitting as a flexible way to estimate
high-resolution column density maps. Our goal is to evaluate the
accuracy of this procedure and to determine whether it is a viable
alternative for making these maps.
}
{
The new method consists of model maps of column density (or intensity
at a reference wavelength) and colour temperature. The model is fitted
using Markov chain Monte Carlo (MCMC) methods, comparing model
predictions with observations at their native resolution. We analyse
simulated surface brightness maps and compare its accuracy with method
B and the results that would be obtained using high-resolution
observations without noise. 
}
{
The new method is able to produce reliable column density estimates at
a resolution significantly higher than the lowest resolution of the 
input maps. Compared to method B, it is relatively resilient against
the effects of noise. The method is computationally more demanding, but
is feasible even in the analysis of large Herschel maps.
}
{
The proposed empirical modelling method E is demonstrated to be a good
alternative for calculating high-resolution column density
maps, even with considerable super-resolution.  Both methods E and B
include the potential for further improvements, e.g., in the form of
better a priori constraints.
}

\maketitle
%

\section{Introduction} \label{sect:intro}

Sub-millimetre and millimetre dust emission data are one of the main
tools in the study of dense interstellar clouds \citep{Motte1998,
Andre2000}. The large surveys of Herschel have provided data on many
nearby molecular clouds \citep{Andre2010} and on the Galactic plane
\citep{Molinari2010}. Dust acts as a tracer of cloud mass, but also the
intrinsic dust properties are actively investigates because they are
expected to reflect the general evolution of the clouds
\citep{Ossenkopf1994,Stepnik2003, Meny2007, Compiegne2011}. The
variations of dust properties can also be an important source of
uncertainty in the column density estimates, but this problem is beyond
the scope of the present paper. 

The interpretation of dust emission data is affected by problems
associated with the effects of noise and temperature variations. The
noise is particularly problematic for attempts to simultaneously
determine both the dust temperature and the dust emissivity spectral
index $\beta$ \citep{Shetty2009b, Juvela2012_chi2}. Therefore, most
estimates of cloud masses are derived by assuming a constant value of
$\beta$. The mass estimates are always affected by temperature
variations, along the line-of-sight or otherwise within the instrument
beam. This tends to lead to underestimation of cloud masses
\citep{Evans2001, StamatellosWhitworth2003, Shetty2009a, Malinen2011,
Juvela2012_TB, YsardJuvela2012}. The effect could be estimated if the
temperature structure of the source is precisely known. In practice, 
detailed modelling is needed but the results still remain uncertain
because the real structure of the clouds and the heating by external
and internal radiation sources are not accurately known
\citep{Juvela2013_PAL}. 

The calculation of column densities requires knowledge of dust
temperature and the intensity of dust emission.  In this paper we
ignore the problems of line-of-sight temperature variations and 
naively assume that the {\em colour} temperature estimated from the
observed intensities can be used to calculate column density. This
could be called an ``apparent column density'' to separate it from the
true column density of the source. In the following, we also use the
terms temperature and column density when actually referring to colour
temperature and the apparent column density.

Colour temperature can be estimated only with multiwavelength
observations where the shape of the observed spectrum is sensitive to
temperature. This is the case, for example, for Herschel, which
carried out photometric observations from 70\,$\mu$m to 500\,$\mu$m.
At the shortest wavelengths, the intensity can be significantly
affected by emission from transiently heated small grains
\citep{Li2001, Compiegne2011}. Therefore, it is safer to base the
estimates of the temperature and column density of large grains on
data between 160\,$\mu$m and 500\,$\mu$m. However, the spatial
resolution of these observations varies from $\sim$12$\arcsec$ at
160\,$\mu$m to $\sim$36$\arcsec$ at 500\,$\mu$m. The usual procedure
is to convolve surface brightness maps to the lowest common
resolution, fit the spectrum in each pixel with a modified black body
law, and use the fitted intensity and colour temperature to calculate
column density estimates at the same resolution.
To obtain a column density map of higher spatial resolution, one
could proceed by directly deconvolving some of the maps
\citep[e.g.][]{Orieux2012}. However, the procedure is sensitive to
noise and to the knowledge of the beam shapes. Therefore, the best
results may not be obtained by carrying out the deconvolution for each
channel independently.

\cite{Juvela2013_PAL} compared two methods that can be used to derive
the column density estimates at a higher resolution. The goal was to
use Herschel data at 160\,$\mu$m--500\,$\mu$m to calculate column
density maps at 18$\arcsec$ resolution. Method A combines
low-resolution colour temperature estimates with the intensity
information of the 250\,$\mu$m observations
\citep{Juvela2012_GCC_III}. In method B, first proposed by
\citet{Palmeirim2013}, column density map is combined from estimates
calculated using different wavelength ranges, the shortest wavelengths
providing the information on the highest spatial frequencies. Method B
was found to be better justified and generally quite reliable, while
method A performed better in cases of a low signal-to-noise ratio (S/N).
The most reliable estimates (and best resilience against noise),
especially in cases of simple source geometry, was obtained by
carrying out radiative transfer modelling of observations. This last
option, however, is also the most complex and most time-consuming
alternative.

In this paper we continue this examination, examining a new method for
calculating high-resolution column density maps. The basic idea
is to set up a model that consists of high-resolution maps of column
density or intensity at some reference wavelength, colour temperature,
and potentially also the spectral index of the dust emission. The
model maps are determined by an optimisation that compares the model
predictions with each of the observed surface brightness maps. Each
comparison is carried out using the native resolution of the observed
maps. In principle, the method is capable of reaching
super-resolution. For comparison with \citet{Juvela2013_PAL}, we set
as the goal column density maps at the same 18$\arcsec$ resolution. We
refer to the method as ``empirical modelling'' to recall that it does
not contain a physical model of the source, but only seeks to describe
the properties of the observed intensities. Therefore, it is subject
to all the errors resulting from temperature variations and from dust
property variations as discussed above.

The content of the paper is the following. In Sect.~\ref{sect:methods}
we present the details of the proposed new method and recount the main
points of method B, which is used as the reference against which the
performance of the new method is compared. In Sect.~\ref{sect:simu} we
described the simulated observations that are then analysed with both
methods in Sect.~\ref{sect:results}. The results are discussed in
Sect.~\ref{sect:discussion} before the final conclusions are presented
in Sect.~\ref{sect:conclusions}.

\section{Methods} \label{sect:methods}

In this section we describe the two methods used to derive column
density maps at a higher resolution than the lowest resolution of the
input surface brightness maps. We refer specifically to the
wavelengths and resolutions of Herschel. However, both methods can be
directly applied to any similar multiwavelength data.

The column density estimates are based on the assumption that observed
intensities follow a modified black body law
\begin{equation}
I_{\nu} = B_{\nu}(T) (1-e^{-\tau}) \approx B_{\nu}(T) \tau =
B_{\nu}(T) \kappa N,
\label{eq:colden}
\end{equation}
where $I_{\nu}$ are the observed intensities, $B_{\nu}(T)$ is the
Planck law for temperature $T$, and $\tau$ is the optical depth.  The
shape of the observed spectrum fixes the colour temperature, and the
optical depth can be solved from the above equation. The optical depth
can be trivially scaled to column density if $\kappa$, the dust
emission cross section per, e.g., hydrogen atom is known. In this
paper we are not interested in the value of $\kappa$; that is to say,
the tests are actually carried out using optical depth $\tau$. One
could equally assume that the exact value of $\kappa$ is known.

\subsection{Method B} \label{sect:MethodB}

In \cite{Palmeirim2013} a higher resolution column density map was
obtained by combining estimates derived using different subsets of 
Herschel wavelengths 160, 250, 350, and 500\,$\mu$m.  One calculates
column density maps $N(250)$, $N(350)$, and $N(500)$ that are based on
surface brightness maps up to the quoted wavelength, all convolved to
the resolution of the quoted wavelength. When the resulting maps are
convolved to a lower resolution, one obtains estimates for the new
column density features that appear between the lower and the higher
resolutions. We use the notation $N(\lambda_1 \rightarrow \lambda_2)$
to denote a column density map that is derived from data at
wavelengths $\lambda \le \lambda_1$ and then convolved to the
resolution of observations at wavelength $\lambda_2$.
The final column density map is obtained as a combination
\begin{eqnarray}
N & = & N(500) +  \left[ N(350) - N(350\rightarrow 500) \right] 
 \nonumber
\\
  &   & +  \left[ N(250) - N(250\rightarrow 350) \right].
\label{eq:palmeirim}           
\end{eqnarray}
where $N(500)$ is the best estimate of column density at low resolution, and
the other terms add information on structures that become visible at
progressively higher spatial resolution. Thus, Eq.~\ref{eq:palmeirim}
in principle provides a column density map at the resolution of the
250\,$\mu$m observations, $\sim 18 \arcsec$.

The estimates $N(250)$, $N(350)$, and $N(500)$ will be different not
only because of the different resolution but also because of the noise,
and bias will be different for different wavelength ranges
\citep{Shetty2009a, Shetty2009b, Malinen2011}. The biases are related
to the temperature distribution of the source, which are present in
real data,
although not in our simulations. In particular, without data at long
wavelengths, one will be relatively insensitive to very cold dust. Of
course, if the estimates were identical, one could directly use the
$N(250)$ map. With Eq.~\ref{eq:palmeirim}, one can include all the
data, although the correction terms $\left[ N(350) - N(350\rightarrow
500) \right]$ and $\left[ N(250) - N(250\rightarrow 350) \right]$
(i.e., estimates of small-scale structures) will be progressively more
insensitive to cold emission.

\subsection{Empirical modelling} \label{sect:empirical}

The principle of the proposed ``empirical modelling'' method (in the
following, method E) is simple. One constructs a model where the free
parameters correspond to maps of optical depth (or directly column
density), temperature, and spectral index. In our implementation the
actual parameters are the 250\,$\mu$m intensity, temperature, and the
spectral index. The 250\,$\mu$m optical depth and the column density
are obtained from Eq.~\ref{eq:colden}. In this paper, we assume a
constant value for the spectral index. The pixel size of these maps is
selected to be smaller than the target resolution, i.e., the
resolution of the estimated column density maps. The model provides
high-resolution predictions of the surface brightness in all observed
bands. The model (consisting of the pixel values in maps of
250\,$\mu$m intensity, temperature, and possibly spectral index) is
optimised by comparing observations with model predictions that are
convolved down to the resolution of the observations.  If some input
data had much higher resolution, they could be compared at some lower
resolution that is still well sampled by the pixel size of the model
maps (and is preferably better than the target resolution). Assuming
that the observational errors follow normal distribution, we minimise
the $\chi^2$ value
\begin{equation}
\chi^2 = \sum_{i}^{N_{\nu}} \sum_{j}^{N_{\rm pix}}
       (I_{i,j} - (M_{i} *  B_{i})_{j})^2,
\end{equation}
where $I_{i,j}$ is the observed intensity for frequency $i$ and pixel
$j$, $M_{i}$ is the model prediction at frequency $i$, $B_{i}$
represents the telescope beam, and the asterisk denotes the
convolution operation. We use the method to analyse simulated
Herschel observations between 160\,$\mu$m and 500\,$\mu$m, and thus the
number of observed frequencies is $N_{\nu}$=4. The number of pixels in
our model maps is $N_{\rm pix}=128 \times 128$ (see
Sect.~\ref{sect:simu}).

The resolution of the 160\,$\mu$m map is taken to be 12.0$\arcsec$, and
the pixel size of the model maps is set to 4.0$\arcsec$. The pixel
size of the model maps is independent of the pixel size of the
observed maps. The only constraint is that the pixelisation should be
fine enough to sample the column density at the target
resolution. If the pixels are very small, the values of individual
model pixels are no longer well constrained by the data, and although
the model predictions match the observations at low resolution, there
could be large oscillations on smaller scales. The oscillations 
should be suppressed with some additional regularisation, especially
if the goal is to reach super-resolution. We include weak
regularisation by adding a penalty to the difference between the model
prediction in a single pixel compared to the convolved prediction at
the observed resolution, $\Delta ln \,p = -[(S_i-<S>)/\delta S]^2$.
The penalty is applied to two bands, 160\,$\mu$m and 250\,$\mu$m, to
constrain both temperature and column density variations. The penalty
is calculated with $\delta S$=1.0 (in the units of the simulated
surface brightness, e.g. Fig~\ref{fig:model_maps}) and, combined with
the small pixel size, turns out to have little effect on the solution.

We solve the model parameters with Markov chain Monte Carlo (MCMC)
calculations that also provide full posterior probability
distributions for the solved 250\,$\mu$m intensity $I_{250}$,
temperature, and for spectral index if that were included as a free
parameter. The probability distribution of $\tau(250\mu{\rm m})$ can
be computed by registering the values of the ratio
$I_{250}$/$B_{250}(T)$. We use flat priors that restrict the
temperature values to the range of 6--30\,K and force the solution to
have non-negative column density. The calculations are quite
time-consuming because, on each step of the MCMC chain, several maps
of model-predicted surface brightness need to be convolved to the
observed resolutions. In the case of the simulations, we can check
convergence (proper burn-in) using not only the changes of the
$\chi^2$ values but also the absolute values. The reduced $\chi^2$
values were below 2.0 before the final runs where the model estimates
were obtained as averages over 50000 MCMC steps.

\section{Simulated observations} \label{sect:simu}

Our goal is to estimate, based on given observed intensities, how good
the methods presented in Sect.~\ref{sect:methods} are in recovering
high-resolution estimates of the apparent column density. We are
interested in the relative performance at different noise levels, but
differences between the apparent and the real column density are
beyond the scope of this study (see Sect.~\ref{sect:intro}).
Therefore, we can use simple simulations, and the methods can be
evaluated entirely based on their ability to reproduce the results
that would be obtained with noiseless observations at a given
resolution.

\begin{figure}
\centering
\includegraphics[width=8.7cm]{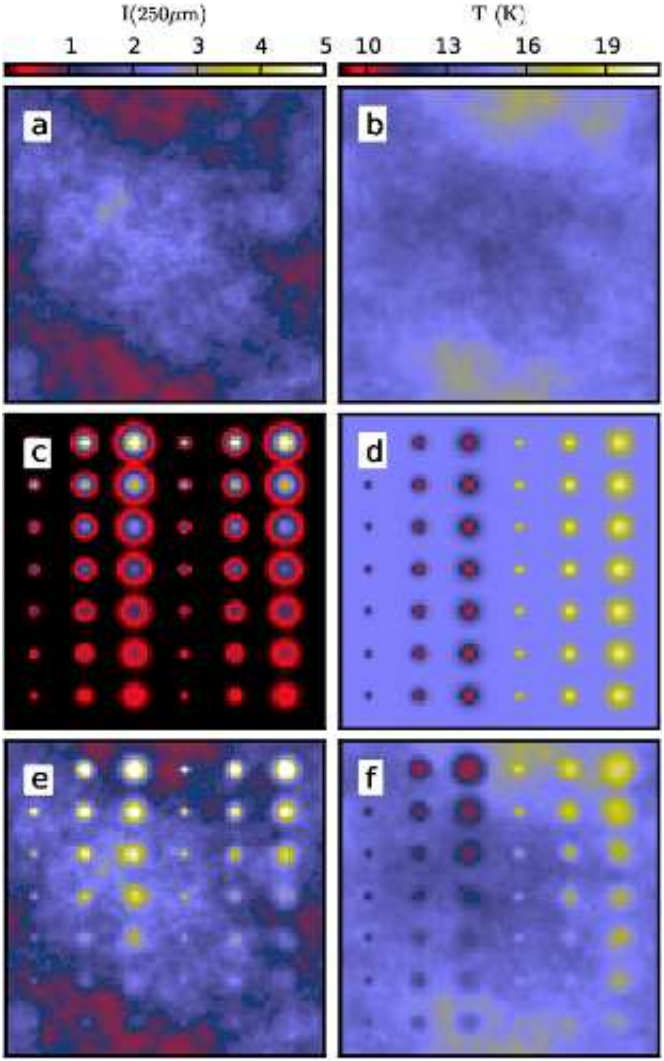}
\caption{
Simulated data used in the tests. The 250\,$\mu$m surface brightness
(left frames, arbitrary units) and the temperature (right frames) are
shown for the three basic test cases.  Shown are the maps for test
cases with diffuse field (frames a-b), compact sources with radial
temperature variations (frames c-d), and isothermal compact sources on
diffuse background (frames e-f).
}
\label{fig:model_maps}
\end{figure}

We simulated three sets of surface brightness maps, each with
different amounts of noise. The spectral index is fixed to a constant
value of 1.8. The first set of maps represents a general fluctuating
intensity field. Data are generated by scaling the amplitudes of
the Fourier elements of 250\,$\mu$m surface brightness and dust
temperature as $\sim k^{-1.5}$ where $k$ is the spatial frequency.
After randomising the phases and transforming the data back to real
space, suitable scaling results in the $I_{250}$ maps shown in the
uppermost frames of Fig.~\ref{fig:model_maps}. 

The second set of maps represents a series of compact isothermal
sources with different sizes and intensities. The radial column
density profile of each source is Gaussian. The temperatures are fixed
to 10\,K for one half of the sources and to 20\,K for the second half. For
each temperature, there are three columns of sources with FWHM sizes
10.0, 20.0, and 30.0 arcseconds (see Fig.~\ref{fig:model_maps}). The
rows of sources correspond to changes in the 250\,$\mu$m peak
intensity by a factor of 1.5. We examine similar sources separately 
with radial temperature variations. The temperatures follow similar
Gaussian profiles to the column density. Instead of being constant
10\,K or 20\,K, the temperature changes from 15\,K in the outside to
10\,K or 20\,K at the centre (Fig.~\ref{fig:model_maps} c-d).

The final set of test maps is the combination of the isothermal
sources and a diffuse background with variations in both intensity and
temperature (Fig.~\ref{fig:model_maps} e-f). The intensity maps are
obtained by adding together the intensities of the previous diffuse
background maps and isothermal sources. Figure~\ref{fig:model_maps}
shows the colour temperature estimated from those intensities.
However, because of the addition of modified black body spectra at two
different temperatures, the resulting spectra are in fact no longer
pure modified black bodies.

We add Gaussian noise to each of the simulated surface brightness
maps. The default noise values are 0.15, 0.05, 0.022, and 0.02 units
at 160\,$\mu$m, 250\,$\mu$m, 350\,$\mu$m, and 500\,$\mu$m. These are
the noise values at the resolution of the simulated observations,
12.0, 18.3, 24.9, and 36.3 arcsec, respectively. The units are
arbitrary but correspond to the surface brightness units in our
simulations (e.g., in Fig.~\ref{fig:model_maps}). In real Herschel
observations, a cold high column density cloud can produce a surface
brightness of $I(250\mu{\rm m}\sim$100\,MJy\,sr$^{-1}$, while the
typical noise value is $\sim$2\,MJy\,sr$^{-1}$, giving a S/N of
$\sim$50 \citep[see e.g.][]{Juvela2012_GCC_III, Kirk2013}. In our
simulations a typical S/N ratio with the default noise values is
similarly $\sim$2.5:0.05=50 (in model B the signal course to zero
outside the sources).  In the following section, we examine cases
where the noise is scaled with a factor $k_{\rm n}=0.1-3.0$ relative
to the default values given above.

\section{Results} \label{sect:results}

\subsection{Diffuse field}

\begin{figure}
\centering
\includegraphics[width=8.7cm]{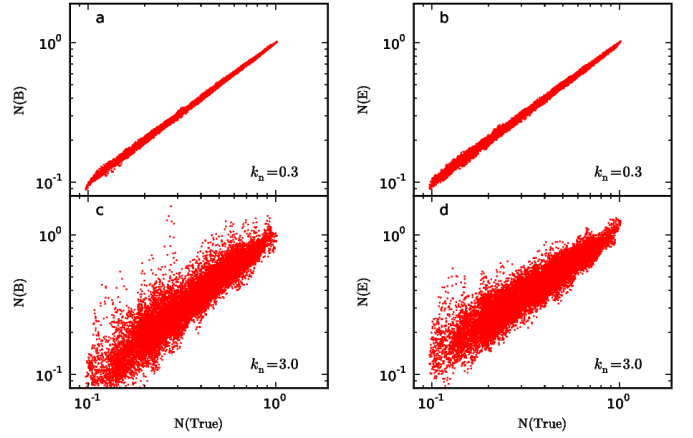}
\caption{
Correlations between the column densities estimated with methods B and
E and the column densities derived from noiseless data with a uniform
resolution of 18.3$\arcsec$ at all wavelengths
160\,$\mu$m--500\,$\mu$m. The data correspond to a diffuse field with
column density and temperature fluctuations.
}
\label{fig:scatterplot_BG}
\end{figure}

The first tests were done with a diffuse field that has a complex
structure in both intensity and temperature and, consequently, in
column density (see Fig.~\ref{fig:model_maps}a-b).
Figure~\ref{fig:scatterplot_BG} compares the column density estimates
that are derived with methods B and E. As in all the following tests,
the observations consist of data at 160\,$\mu$m--500\,$\mu$m.

At the noise level $k_{\rm n}=0.3$, method B gives slightly lower
scatter than method E (0.007 vs. 0.009 units). Method E values are to
some extent affected by Monte Carlo noise because the presented values
are averaged over only 50000 MCMC steps. Both methods recover the true
column density without noticeable bias. For the higher noise case,
$k_{\rm n}$=3.0, the situation is reversed and the results of method E
have an rms error of 0.073 units compared to 0.083 units for method B.
This is to be expected because, by also using column density estimates
calculated from fewer frequency bands, method B should be more
sensitive to noise. Even with $k_{\rm n}$=3.0, the 250\,$\mu$m S/N
is still above 13 everywhere in the maps. 

\subsection{Compact sources}

\begin{figure}
\centering
\includegraphics[width=8.7cm]{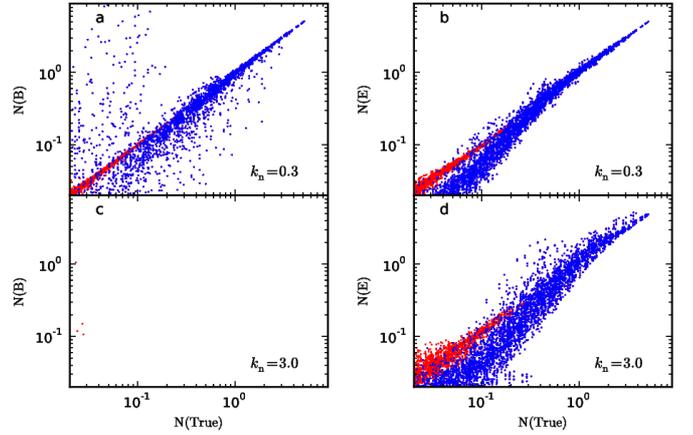}
\caption{
Pixel-by-pixel correlations between the column density estimates of
methods B and E and the column densities derived from noiseless data
with a uniform resolution of 18.3$\arcsec$. The observations
correspond to isothermal sources at 10\,K (blue points) and 20\,K (red
points).
}
\label{fig:scatterplot_PS}
\end{figure}

We examine isothermal compact sources with zero background next. 
Figure~\ref{fig:scatterplot_PS} shows the overall correlations between
the true column density and the estimates derived with methods B and
E. Unlike the previous case, the signal goes to zero outside the
compact sources; i.e., a large fraction of the maps has S/N below one.
In Fig.~\ref{fig:scatterplot_PS} we also indicate the pixels in the
$T=10$\,K sources and in the $T=20$\,K sources with different colours.
These differ especially regarding the 160\,$\mu$m data where the
colder temperature leads to significantly lower signal. For $k_{\rm
n}=0.3$, the two methods give comparable results for the warm sources,
while for the colder sources method B results in a much larger
scatter. The MCMC calculations were started with the observed
250\,$\mu$m surface brightness and a constant temperature of 15\,K.
For the 20\,K sources the calculations have converged to a good
solution. However, for the 10\,K sources, because of the lower S/N,
the values are affected by the temperature priors. The
temperature prior forces temperatures in the range between 6\,K and
30\,K. The 6\,K limit and the large temperature uncertainties bias the
MCMC estimates so that the mean values from the MCMC calculation are
mainly above the correct value of 10\,K. Consequently, the recovered
column density values are systematically too low, and this produces the
bias seen in Fig.~\ref{fig:scatterplot_PS}. It is for the same pixels
that method B exhibits a very large scatter. The same trend extends to
the case with $k_{\rm n}=3.0$. However, method B is no longer able to
produce reasonable estimates even for the 20\,K sources.

\begin{figure*}
\centering
\includegraphics[width=15.0cm]{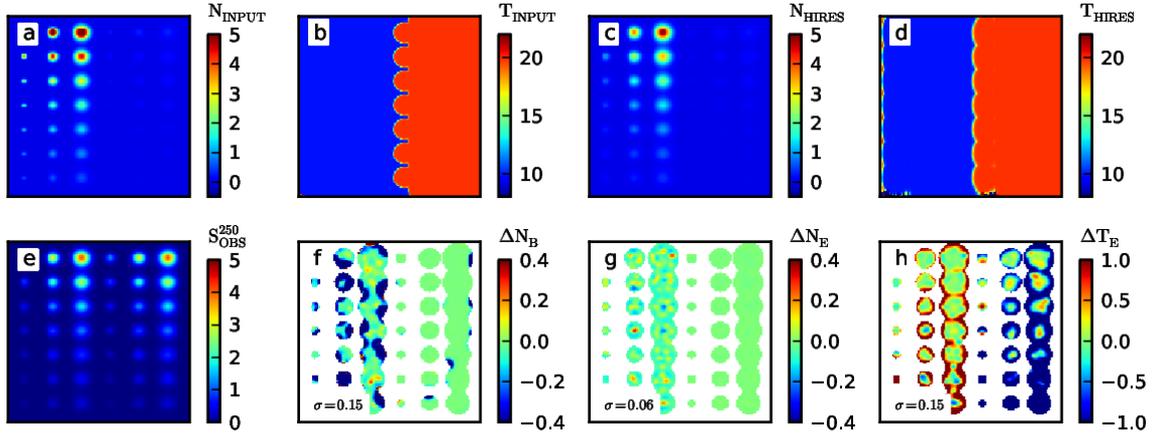}
\caption{
Results for the test with isothermal compact sources with noise
$k_{\rm n}=0.3$. Shown are the input model of column density (frame a)
and temperature (frame b), the corresponding parameters extracted from
noiseless data at 18.3$\arcsec$ resolution (frames c-d), the observed
250\,$\mu$m surface brightness (frame e), the absolute errors of the
column density estimates of method B (frame f), and the absolute
errors of the column density and temperature determined with method E
(frames g and h). In the last three frames, the rms error of the
corresponding parameter is given in the figure.
}
\label{fig:comparison_PS}
\end{figure*}

Figure~\ref{fig:comparison_PS} shows the situation in map form for the
$k_{\rm n}=0.3$ case. In this figure, we have masked in the result
maps all regions where the 250\,$\mu$m surface brightness drops below
the 1-$\sigma$ noise value. Within the unmasked area, the rms error of
method B column density estimates is 0.15, compared to 0.06 for method
E. As the value of $k_{\rm n}$ is increased, the difference between
the methods increases rapidly. As mentioned above, part of this may be
attributed to the fact that, even in the absence of observations of
significant S/N, the temperature of method E remains constrained
by the priors.

We separately tested the accuracy with which the parameters of the
radial column density profiles of the sources were recovered. Based on
the known locations of the Gaussian sources, we fitted the column
density estimates with two-dimensional Gaussian surfaces. The values
were compared not to the input values of the simulations but to the
values that would be recovered in the absence of observational noise,
using 18.3$\arcsec$ resolution data at all four frequencies. Each
fit was carried out with data up to a distance equal to FWHM from the
source centre.

The injected sources have FWHM values of 10.0, 20.0, and 30.0
arcseconds. Because of the convolution with the 18.3$\arcsec$ beam,
the extracted FWHM values should be equal to 20.9$\arcsec$,
27.1$\arcsec$, and 35.1$\arcsec$, respectively. The peak values are
similarly lower than in the injected Gaussians but are compared
directly with the results from idealised observations at the
18.3$\arcsec$ resolution.

\begin{figure}
\centering
\includegraphics[width=8.7cm]{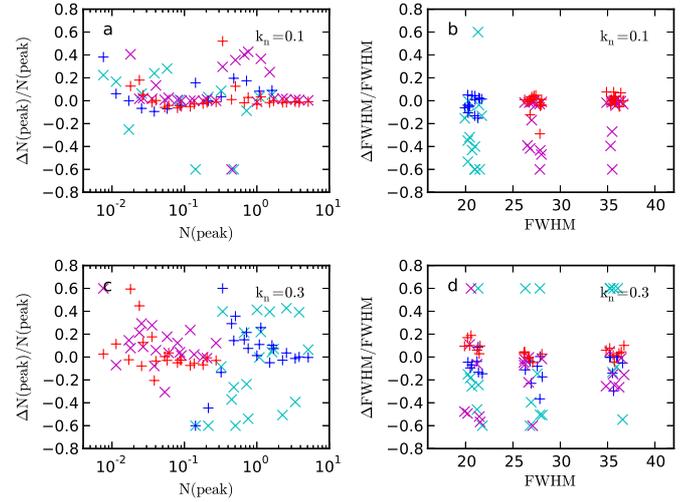}
\caption{
The accuracy of the recovered peak column density (left frames) and
FWHM (right frames) values of the compact isothermal sources. The
plots show the relative errors for method B (crosses) and method E
(plus signs) relative to idealised observations at 18.3$\arcsec$
resolution. The colours are used to separate 10\,K and 20\,K sources,
blue and red for method E and cyan and magenta for method B,
respectively. The upper frames correspond to a noise level of $k_{\rm
n}=0.1$, the lower frames to $k_{\rm n}=0.3$. Values outside the range
of [-0.6, 0.6] are plotted at those boundaries. In the FWHM plots, to
reduce the overlap of the symbols, we have added jitter in the x-axis
values.
}
\label{fig:plot_gaussian_fits_PS}
\end{figure}

Figure~\ref{fig:plot_gaussian_fits_PS} illustrates the accuracy of the
recovered peak column density and FWHM values relative to these
reference values. The noise corresponds to the low noise level of
$k_{\rm n}=0.1$ where approximately correct values are still recovered
for most of the sources. The accuracy is again better for method E, and
this is particularly noticeable for the estimated FWHM values. 

When the noise level is raised to $k_{\rm n}=0.3$
(Fig.~\ref{fig:plot_gaussian_fits_PS}, lower frames), some of the
lower intensity sources are lost and many of the FWHM estimates of
method B are already off by more than 60\%. Based on
Fig.~\ref{fig:scatterplot_PS}, this already was to be expected. The
difference is partly because method E simultaneously uses all
wavelengths to constrain the solution. However, equally important is
the technical detail that method E calculations include priors that
prevent the column density estimates from exploding in the
noise-dominated outer regions of the sources.

\begin{figure}
\centering
\includegraphics[width=8.7cm]{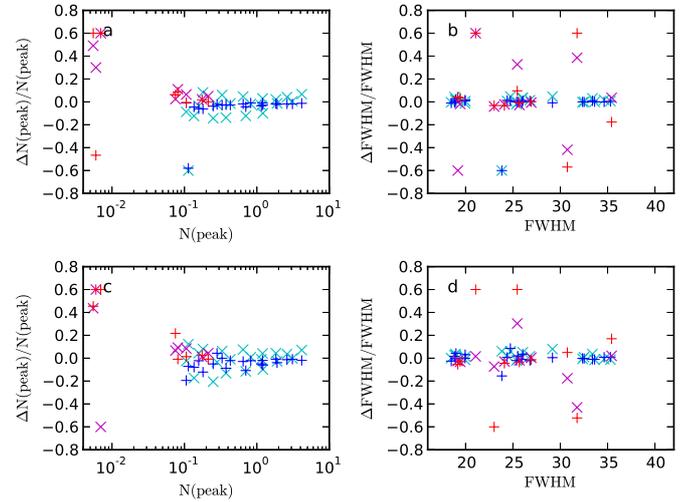}
\caption{
The accuracy of the recovered peak column density (left frames) and
FWHM (right frames) values for compact sources on fluctuating
background. The plots show the relative errors for method B (crosses)
and method E (plus signs) relative to idealised observations at
18.3$\arcsec$ resolution.  The 10\,K and the 20\,K sources are
respectively shown as blue and red symbols for method E and as cyan
and magenta symbols for method B. The upper frames correspond to a
noise level of $k_{\rm n}=0.1$, the lower frames to $k_{\rm n}=0.3$.
Values outside the range of [-0.6, 0.6] are plotted at those
boundaries. 
}
\label{fig:plot_gaussian_fits}
\end{figure}

When the isothermal sources are seen against a fluctuating diffuse
background, we attempt to recover the source parameters by fitting a
two-dimensional Gaussian surface plus a constant background. For low
values of $k_{\rm n}$, the observations of the sources are no longer
noise-limited, and all sources cannot be detected even in the noiseless
maps. Therefore, in Fig.~\ref{fig:plot_gaussian_fits}, we only compare
the predictions of methods B and E when the noiseless 18.3$\arcsec$
resolution observations recover both the peak column density and the
FWHM extent of a source to a relative accuracy better than 50\%. This
removes most sources with the lowest peak column densities but also
some of the stronger sources that are seen towards a high background. 

Compared to Fig.~\ref{fig:plot_gaussian_fits_PS}, the scatter in the
parameters recovered by methods B and E is decreased because of this
filtering of sources. Method E appears to be more accurate in
extracting the values of peak column density while, unlike in
Fig.~\ref{fig:plot_gaussian_fits}, the difference is small on the FWHM
estimates for $k_{\rm n}=0.3$. The rms errors are very similar because
these values are determined mostly by the outliers (i.e. points with
relative errors above $\sim$0.2) that are equally frequent for both
methods.

The final test involves compact sources with radial temperature
profiles. The background has a constant temperature of 15\,K, and the
temperature either decreases to 10\,K or increases to 20\,K at the
centre of a source (see Fig.~\ref{fig:model_maps}). Figures
~\ref{fig:scatterplot_PS} and ~\ref{fig:plot_gaussian_fits_PT} again
show the pixel-to-pixel correlations and the accuracy of the recovered
Gaussian parameters in relation to idealised observations. Comparing
the result to Fig.~\ref{fig:scatterplot_PS}, the noise is actually
lower, mainly because the observed temperature of the non-isothermal
sources is always above 10\,K. For the same reason, there is no bias
in method E results where the column density of the colder sources
would be systematically underestimated. In contrast, the relative
error in the column density of the 10--15\,K sources appears to be
positive. This is not obvious for method B (possibly because of the
large scatter) but is clear for method E. It could be related to the
fact that, because of the higher S/N, the central part of each source
has a large weight in determining the overall solution. This could be
enhanced by the smoothing of the model maps but does not appear likely
because of the weakness of that regularisation. In the higher S/N
case, method B performs well providing actually more accurate
estimates of $N$ for the highest column density sources, again partly
be caused of the Monte Carlo noise of method E calculations.

\begin{figure}
\centering
\includegraphics[width=8.7cm]{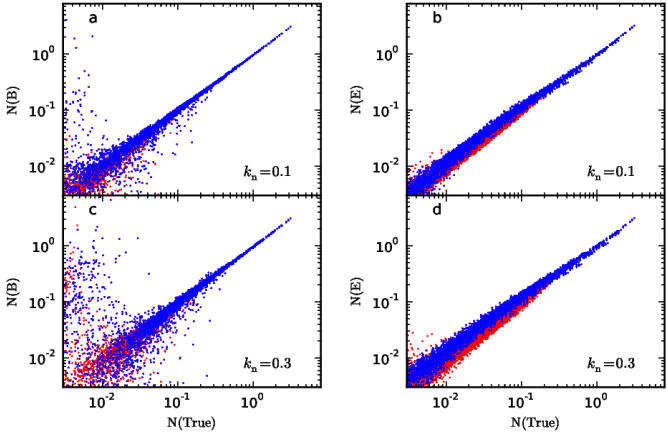}
\caption{
Correlations between method B column density predictions (left frames)
and method E column density predictions (right frames) with the values
derived from noiseless data with a uniform resolution of
18.3$\arcsec$. The model consists of compact sources with radially
increasing or decreasing temperature. The noise levels $k_{\rm n}$ are
indicated in the frames. The colours indicate the pixels in the warm
sources (15--20\,K, red points) and the cold sources (10--15\,K, blue
points).
}
\label{fig:scatterplot_PT}
\end{figure}

\begin{figure}
\centering
\includegraphics[width=8.7cm]{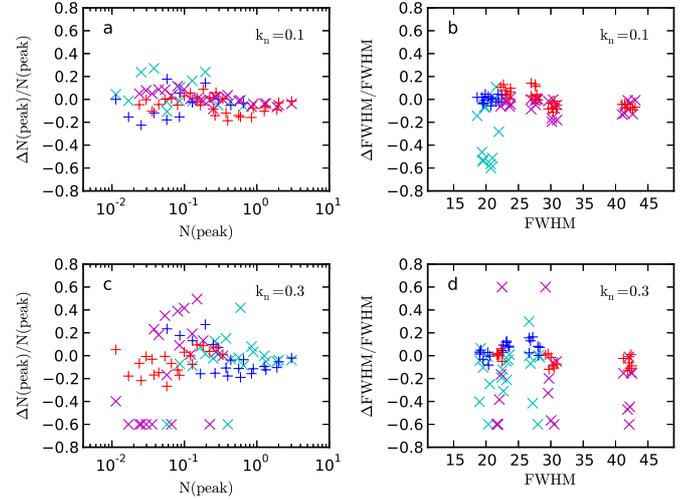}
\caption{
Corresponding to the models in Fig.~\ref{fig:scatterplot_PT}, the
accuracy of the recovered peak column densities (left frames) and FWHM
values (right frames). The relative errors of method B (cyan and
magenta crosses for cold and warm sources, respectively) and method E
(blue and red plus signs for cold and warm sources, respectively) are
shown relative to idealised observations at 18.3$\arcsec$ resolution.
The noise levels $k_{\rm n}$ are indicated in the frames. Values with
relative errors outside the range of [-0.6, 0.6] are plotted at those
boundaries.  To reduce overlap between plot symbols, some jitter was
added to the x-coordinate values of the FWHM plots.
}
\label{fig:plot_gaussian_fits_PT}
\end{figure}

\section{Discussion} \label{sect:discussion}

We have compared two methods, methods B and E, that can be used to
derive high-resolution column density maps from multifrequency
observations of dust emission. This study is an extension of the paper
by \citet{Juvela2013_PAL} where some methods already were discussed,
including the method B. In this paper we have proposed method E, the
``empirical modelling'', as a viable alternative for the task. In this
method high-resolution model images of column density (or intensity at
a reference wavelength) and temperature are, via convolution, matched
to observations. This has the advantage that all wavelengths are used
simultaneously to constrain the solution. The calculations were
completed with MCMC methods, using flat priors that allowed all
positive values of column density and all temperatures in the range
6\,K$<T<$30\,K. 

We carried out tests with simulated observations corresponding to
Herschel data at wavelengths between 160\,$\mu$m and 500\,$\mu$m. The
tests conducted with fields of diffuse surface brightness and with
fields containing compact objects both showed that the method E
provides generally more accurate results. However, the differences
were mostly relatively small and became significant only when dealing with
data with low S/Ns. The advantages of method E result partly
from the basic nature of the method, i.e., the simultaneous fit of
data at all wavelengths. In method B, the highest spatial frequencies
are estimated based only on the 160\,$\mu$m and 250\,$\mu$m bands.
This makes the method sensitive to observational noise in those
channels, particularly in the case of cold dust emission for
which the S/N of the 160\,$\mu$m band can be low,
both in our simulations and in real cold dust clouds.

The Bayesian implementation of method E is equally important. Although
the priors were flat and had wide ranges of allowed values, they
become important at low signal levels. This was most noticeable in
tests with compact sources on zero background (e.g.,
Fig.~\ref{fig:scatterplot_PS}). In the outer regions of the sources
the S/N falls below one, but in method B, the column
density estimates start to fluctuate strongly already well before
that. In our implementation of method E, the temperature is
constrained to remain above 6\,K. The prior applies to all MCMC
samples. For low S/Ns, the credible region of $T$ becomes very
wide. The constraint $T>6$\,K removes the low-temperature samples, and
this can bias the Bayesian estimate, which is the average of all MCMC
samples, of cold sources. The effect was seen in
Fig.~\ref{fig:scatterplot_PS} where the column density estimates were
biased downwards for most 10\,K, low surface-brightness sources.
However, a limit on $T$ values is essential to avoid very low
temperatures that could, in a strongly non-linear fashion, lead to
very high column densities (as was the case of method B). In 
Fig.~\ref{fig:scatterplot_PS}, a more liberal lower limit would reduce
the bias of the temperature estimates of method E but, as shown by the
corresponding results of method B, the data are clearly not
enough to strongly constrain the results. This is readily seen in
the MCMC error estimates of the temperature.

The radial temperature variations did not cause particular problems
for either method
(Figs.~\ref{fig:scatterplot_PT}-\ref{fig:plot_gaussian_fits_PT});
i.e., the errors were not significantly larger than what would have
been expected based on the tests with isothermal sources.  There was
only a weak indication of a possible bias that may affect column density
estimates in the outer parts of the sources.

Method E compares a high-resolution model with observation through
convolution and is thus essentially a deconvolution method. We defined
the models at 4$\arcsec$ pixels, and the model predictions were
convolved to a 18.3$\arcsec$ resolution for comparison with method B.
The final resolution can be selected freely as long as it is lower
than the resolution of the model itself. However, because
deconvolution amplifies the noise, the solution becomes more uncertain
with higher target resolution, and the solution may need additional
regularisation to avoid unphysical oscillations. So far, the
regularisation consisted of the priors, and a small penalty that was
attached to surface brightness changes between the neighbouring pixels
of the full-resolution model predictions. Because many of the
observations existed at or close to the target resolution, there is
still little need for regularisation as long as the S/N is high.

\begin{figure}
\centering
\includegraphics[width=8.7cm]{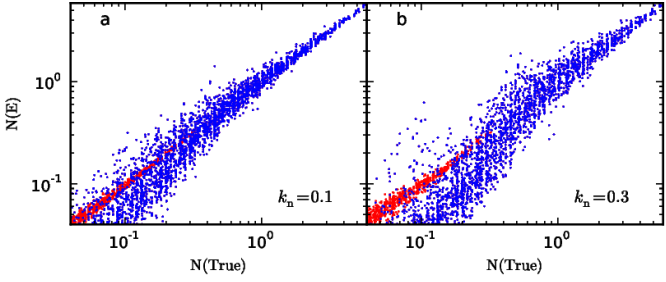}
\caption{
Correlations between the column density estimates of method E and the
corresponding values derived from noiseless data with a uniform
resolution of 12.0$\arcsec$. The noise levels $k_{\rm n}$ are given in
the figure. The blue and the red points correspond, respectively, to
pixels in 10\,K and in 20\,K sources.
}
\label{fig:scatterplot_AS12}
\end{figure}

\begin{figure}
\centering
\includegraphics[width=8.7cm]{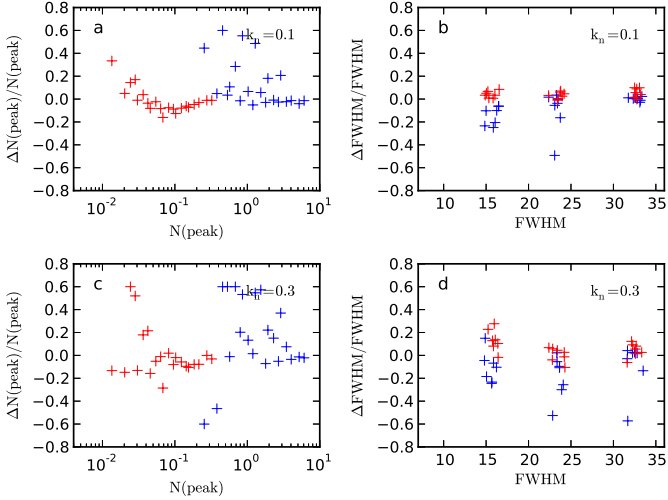}
\caption{
The accuracy of the recovered peak column density (left frames) and
FWHM (right frames) values for compact sources, estimated with method
E at the resolution of 12.0$\arcsec$. The upper frames correspond to a
noise level of $k_{\rm n}=0.1$ and the lower frames to $k_{\rm
n}=0.3$. The blue and the red symbols correspond, respectively, to the
10\,K and 20\,K sources. Values outside the range of [-0.6, 0.6] are
plotted at those boundaries. To reduce the overlap of plot symbols,
some jitter has been added to the x-axis values of the FWHM plots.
}
\label{fig:gaussians_AS12}
\end{figure}

As a final test, we calculated model predictions at 12.0$\arcsec$
resolution. Apart from the change in resolution, the test corresponds
to that of Fig.~\ref{fig:scatterplot_PS}. The results of method E were
compared with parameters that would be obtained with noiseless
observations where the resolution is 12.0$\arcsec$ at all wavelengths.
Figure~\ref{fig:scatterplot_AS12} shows the pixel-to-pixel
correlations. Compared to the lower resolution results in
Fig.~\ref{fig:scatterplot_PS}, the noise is, of course, higher because
the noise per resolution unit is more than 50\% higher. The errors are
still small for the 20\,K data and (as relative errors) for all high
column density pixels. In other words, the peak column densities are
recovered with good accuracy even at 12.0$\arcsec$ resolution.
Figure~\ref{fig:gaussians_AS12} directly compares the accuracy of the
recovered peak column density and FWHM values that are obtained from
fitting the sources with 2D Gaussian surfaces. At $k_{\rm n}$=0.1,
1-$\sigma$ relative errors are 18\% and 11\% for column density and
FWHM, respectively. These values are again raised by a small number of
outliers (these, however, are clipped to the range of $\pm$60\%
relative errors, similar to Fig.~\ref{fig:gaussians_AS12}) and most
errors are below 10\% (69\% and 83\% for column density and FWHM,
respectively).

Based on the conducted tests, we can conclude that method E is a
viable alternative for deriving high-resolution column density maps
from observations, such as those conducted with Herschel. One of the
main downsides of the method is its computational cost that is an
issue, regardless of whether the method is implemented as an MCMC
calculation (as in this paper) or as a direct optimisation problem.
For accurate estimation of the uncertainties, MCMC is the better,
although slower alternative. The main cost is caused by, on each
iteration, the model predictions needing to be convolved to each
resolution of the observations. In the tests in this paper, this meant
that 128$\times$128 pixel surface brightness maps were convolved to
four different resolutions (corresponding to 160\,$\mu$m, 250\,$\mu$m,
350\,$\mu$m, and 500\,$\mu$m measurements. The burn-in and the proper
sampling of the credible region both required some tens of thousand
MCMC steps. This resulted in a Monte Carlo noise that, however, was
not completely negligible at small $k_{\rm n}$ (e.g.,
Fig.~\ref{fig:scatterplot_PS}). The calculations can be sped-up by
using GPUs for the convolutions. Nevertheless, the calculation times
are several hours in wall-clock time, depending somewhat on the
selected initial conditions. We started all runs with a constant
temperature of 15\,K, but the burn-in phase could be shortened by
using, for example, the results of method B as the starting point.
With 4.0$\arcsec$ pixel size, the analysed maps corresponded to an
area of $\sim 8.5\arcmin \times 8.5\arcmin$. This means that the
analysis of one-degree-sized maps could take several days. Although
long, this can still be considered feasible. The convolution makes the
problem non-local; i.e., model parameters depend on each other over
distances defined by the largest beam of observations. However, the
observations could easily be divided into patches that could be
processed in parallel.

In the tests the dust emission spectral index $\beta$ was assumed to
be constant. It is straightforward to include variable $\beta$ in the
model by defining the model through maps of column density,
temperature, and spectral index. Because of the natural
anticorrelation of $T$ and $\beta$ in the modified black body fits,
this would significantly increase the scatter in the estimated
temperatures \citep{Shetty2009b, Juvela2012_chi2, Juvela2013_TB} and
in $N$. Therefore, it might be necessary to constrain the solutions
more strongly, e.g., by effectively determining the spectral index
only at a lower resolution. However, this is not an optimal solution
because the (apparent) spectral index can change even on small scales,
in association with local heating sources \citep{Malinen2012,
Juvela2011_GCC_II}. It would be better to restrict the use of strong
regularisation to regions of low surface brightness. This could be
implemented within the framework of hierarchical models that also
would allow the inclusion of a wide range of other constraints
\citep{Kelly2012, Juvela2013_TB}.

We carried out preliminary tests that showed that, if allowed by the
S/N, method E can be used directly for super-resolution,
reaching scales below those directly accessible to method B.  More
investigations would be needed to quantify the performance and to find
out whether the results could be further improved, e.g., with more
realistic priors. Further work also should be done on the method B. It
may be possible to significantly reduce its sensitivity to noise
simply by using constrained optimisation, e.g., by effectively adopting
similar priors as were already used in method E.

\section{Conclusions} \label{sect:conclusions}

We have proposed ``empirical modelling'' as method E, for estimating
high-resolution column density maps. As such, this conceptually
straightforward method can be used whenever observations of different
resolutions need to be combined. In this paper, the method was applied
to simulated dust emission observations that correspond to different
resolution Herschel measurements at wavelengths between 160\,$\mu$m
and 500\,$\mu$m. The method was compared to another method, called
here method B, which was presented by \cite{Palmeirim2013} and further
discussed in \cite{Juvela2013_PAL}. The tests lead to the following
conclusions.
\begin{itemize}
\item For high S/N data, both methods provide reliable 
column density estimates at the $\sim$18$\arcsec$ resolution of
250\,$\mu$m data.
\item Method B is more sensitive to noise. This
also complicates the analysis of compact objects because of the strong
column density fluctuations towards the outer parts of the sources.
\item Method E is more time-consuming but gives consistently more
reliable estimates. The use of GPU computing makes the method feasible
for analysing large maps.
\item Radial temperature variations has no significant effect on the
accuracy of the column density estimates of either method.
\item With data of sufficient S/N, method E can be
used directly for even higher resolution estimates. This was
demonstrated by estimating column densities at 12$\arcsec$ resolution.
\item Further improvements should be possible in both methods. For
method E, the super-resolution estimates may be improved with better
regularisation. For method B, constrained optimisation can mitigate
the problems encountered at low S/N levels.
\end{itemize}

\begin{acknowledgements}
The authors acknowledge the support of the Academy of Finland grant No. 250741.
\end{acknowledgements}

\bibliography{biblio_v2.0}

\end{document}